\documentclass[conference]{IEEEtran}
\IEEEoverridecommandlockouts
\usepackage{cite}
\usepackage{amsmath,amssymb,amsfonts}
\usepackage{algorithmic}
\usepackage{graphicx}
\usepackage{textcomp}
\usepackage{xcolor}
\usepackage{listings}
\usepackage{listings-rust}
\usepackage{soul}
\usepackage{paralist}

\newcommand{\fig}[1]{Fig.~\ref{fig:#1}}

\newcommand{\s}[1]{Sec.~\ref{sec:#1}}

\newcommand{\fakeparagraph}[1]{\vspace{1mm}\noindent\textbf{#1.}}
\newcommand{\emphparagraph}[1]{\vspace{1mm}\noindent\emph{#1.}}

\def\BibTeX{{\rm B\kern-.05em{\sc i\kern-.025em b}\kern-.08em
    T\kern-.1667em\lower.7ex\hbox{E}\kern-.125emX}}
\begin{document}

\title{FlowUnits: Extending Dataflow for the Edge-to-Cloud Computing Continuum}

\author{
\IEEEauthorblockN{Fabio Chini}
\IEEEauthorblockA{\textit{Politecnico di Milano, Italy}\\
fabio.chini@polimi.it}
\and
\IEEEauthorblockN{Luca De Martini}
\IEEEauthorblockA{\textit{Politecnico di Milano, Italy}\\
luca.demartini@polimi.it}
\and
\IEEEauthorblockN{Alessandro Margara}
\IEEEauthorblockA{\textit{Politecnico di Milano, Italy}\\
alessandro.margara@polimi.it}
\and
\IEEEauthorblockN{Gianpaolo Cugola}
\IEEEauthorblockA{\textit{Politecnico di Milano, Italy}\\
gianpaolo.cugola@polimi.it}
}

\maketitle

\begin{abstract}
This paper introduces FlowUnits, a novel programming and deployment model that extends the traditional dataflow paradigm to address the unique challenges of edge-to-cloud computing environments. While conventional dataflow systems offer significant advantages for large-scale data processing in homogeneous cloud settings, they fall short when deployed across distributed, heterogeneous infrastructures. FlowUnits addresses three critical limitations of current approaches: lack of locality awareness, insufficient resource adaptation, and absence of dynamic update mechanisms. FlowUnits organize processing operators into cohesive, independently manageable components that can be transparently replicated across different regions, efficiently allocated on nodes with appropriate hardware capabilities, and dynamically updated without disrupting ongoing computations. We implement and evaluate the FlowUnits model within Renoir, an existing dataflow system, demonstrating significant improvements in deployment flexibility and resource utilization across the computing continuum. Our approach maintains the simplicity of dataflow while enabling seamless integration of edge and cloud resources into unified data processing pipelines.
\end{abstract}


\section{Introduction}
\label{sec:intro}

In recent years, the dataflow model~\cite{dataflow_model} has become a well-established paradigm for large-scale data analytics in cloud environments~\cite{data_intensive}. Its design, based on Directed Acyclic Graphs (DAGs) of operators that do not share state, enables seamless parallelization and distribution of computations, making them inherently scalable. This approach has been widely adopted in batch and streaming frameworks such as Apache Flink~\cite{flink} and Apache Spark~\cite{spark}, and it is increasingly seen as an enabling technology to build complex data pipelines~\cite{evolution_streaming, tspoon}.

However, dataflow systems rely on strong assumptions about the homogeneity of the underlying deployment platform, which limits their applicability in more heterogeneous environments such as the edge-to-cloud computing continuum~\cite{continuum}. This paradigm is gaining increasing relevance in application domains like smart cities and the Internet of Things (IoT), where computational resources are more distributed and dynamic.
Specifically, we identified three core limitations that prevent or limit the applicability of the dataflow paradigm to computing continuum scenarios.

\emphparagraph{Lack of locality awareness}
Modern dataflow systems are designed to maximize resource utilization across available computing nodes, without considering their physical location. This approach works well in centralized cloud environments, but becomes problematic in geographically distributed scenarios, such as the edge-to-cloud continuum. In these contexts, data is often generated at the edge -- through IoT devices and sensor networks -- but existing dataflow frameworks lack mechanisms to prioritize local processing before offloading tasks to remote cloud nodes. As a result, organizations are forced to deploy ad-hoc solutions, using separate frameworks for edge and cloud processing. This fragmentation increases development complexity, leads to duplicated logic, and complicates system orchestration, ultimately reducing efficiency and scalability.

\emphparagraph{Lack of resource awareness}
Existing dataflow systems lack mechanisms to control on which specific nodes a computation (or part of it) should be executed. Current frameworks assume a homogeneous execution environment and distribute tasks indiscriminately, potentially leading to inefficient execution on underpowered nodes or missing opportunities to leverage hardware accelerators such as GPUs or TPUs. In scenarios spanning edge, fog, and cloud layers, each with distinct computational capabilities, this lack of resource awareness leads to unfeasible or suboptimal deployments, preventing applications from fully exploiting available heterogeneous resources.

\emphparagraph{Lack of dynamic update mechanisms}
Existing dataflow systems are not designed to support dynamic updates to a running computation. This limitation is particularly critical in streaming applications, where processing is continuous and time constraints are stringent. For many current dataflow systems, even small changes to a data-processing pipeline require stopping the entire computation, buffering incoming data, redeploying the updated operators, and restarting execution. This rigid approach not only introduces downtime but also increases development effort, as engineers must manually handle and coordinate updates. In an edge-to-cloud environment, where new nodes may join dynamically or data from specific regions may require different processing logic over time, this lack of flexibility prevents seamless adaptation.

\vspace{1mm}

To address these challenges, this paper introduces a new programming and deployment model that extends the traditional dataflow paradigm to better suit edge-to-cloud environments. The proposed model retains the key advantages of dataflow, such as simple code definition and automatic distribution, while overcoming its limitations in heterogeneous and dynamic infrastructures.
In our model, developers define computations using familiar dataflow abstractions, and specify hardware requirements and desired geographical deployments as annotations.
Based on these annotations, the operators that compose the dataflow graph are organized into units of deployment that we denote as \emph{FlowUnits}.
FlowUnits that perform the same computation are replicated across geographical regions that demand that type of computation.
Within each geographical region, the operators that compose the FlowUnits replicated on that region are deployed only on nodes with hardware capabilities that align with their processing requirements.
Finally, FlowUnits can be dynamically added, removed, and updated without disrupting the rest of the deployment, enabling for fine-grained adaptation of the processing tasks and their deployment over time.

The contributions of this paper are as follows: we present in detail the FlowUnits programming and deployment model, and we demonstrate and evaluate an initial implementation of this model within Renoir~\cite{renoir}, an existing dataflow system.

The paper is organized as follows. \s{background} introduces background concepts and motivates our work. \s{model} presents our proposed model in detail. \s{impl} overviews its implementation within the Renoir dataflow system, and \s{eval} evaluates it. \s{related} surveys related work and \s{conclusions} provides conclusive remarks and discusses future work.
\section{Background and Motivations}
\label{sec:background}

In the dataflow programming model~\cite{dataflow_model}, computation is organized into a directed graph, where vertices represent operators and edges represent the flow of data between operators.
Since operators do not share any state, this model facilitates parallelization and distribution by deploying operators in multiple instances, each processing an independent partition of the input data in parallel with others, on the same or different hosts.
Developers need only express the logic of each computational step with respect to individual data elements, and the runtime manages operator deployment, synchronization, scheduling, and data communication -- arguably the most complex and critical aspects in distributed applications.

Data processing systems have implemented the dataflow model using two orthogonal execution strategies.
Systems such as Apache Spark~\cite{spark} dynamically schedule operator instances across nodes. Communication between instances occurs by saving intermediate results on shared storage, with operators deployed as close as possible to the input data they consume.
Systems such as Apache Flink~\cite{flink} deploy all operator instances before starting the computation. Communication takes place through message passing between instances.
In this paper, we focus on streaming computations and consider the second execution strategy, which incurs lower latency as it avoids the overhead of operator scheduling at runtime.

\begin{figure}[pt]
\centering
\includegraphics[width=.8\columnwidth]{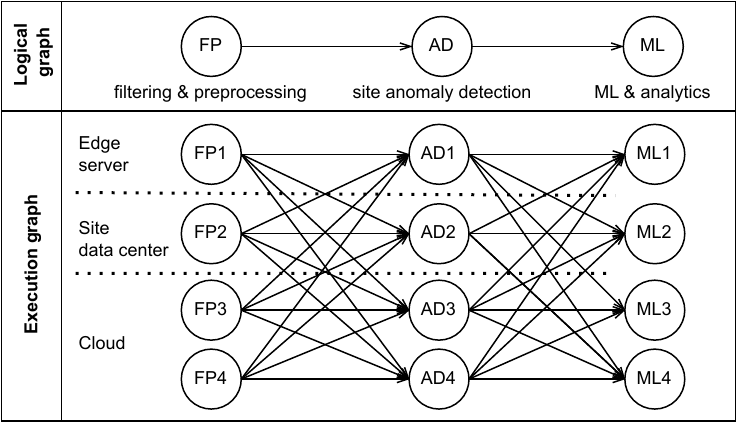}
\caption{Example of dataflow logical and execution graph.}
\label{fig:dataflow_example}
\vspace{-2mm}
\end{figure}

To illustrate these concepts, consider the following example: the Acme company operates multiple production sites and aims to monitor the temperature of each production machine.
The goal is to enable immediate intervention in case of overheating while also aggregating data for analytics.
The upper part of \fig{dataflow_example} shows the logical dataflow graph representing the main steps involved in the process. First, a filtering and preprocessing (\texttt{FP}) step cleans the data coming from individual machines, then an anomaly detection (\texttt{AD}) step is performed considering the data collected from all machines within a given site, and finally a machine learning (\texttt{ML}) model extracts deeper insights from the data coming from all sites.

Let us assume that Acme's compute infrastructure consists of three layers: an edge server connected to each machine, a small data center within each site, and a cloud environment.
To exploit data locality, Acme engineers could perform the \texttt{FP} step for a given machine only within the edge server connected to that machine. Each site's data center could be responsible for the \texttt{AD} step for its site. The cloud platform could then perform the \texttt{ML} step, which requires (filtered) data from all sites.
However, most current dataflow systems operate under the assumption that all nodes are uniform and co-located, and instantiate copies of each operator within each node, typically deploying one copy for each CPU core.
The bottom part of \fig{dataflow_example} illustrates this concept by showing a possible execution graph for the Acme use case. For ease of illustration, \fig{dataflow_example} shows a single edge server with 1 CPU core, a single site data center with 1 CPU core, and a cloud platform with 2 CPU cores. Real-world examples would include multiple edge servers and sites, each providing many CPU cores.
A default execution strategy would instantiate a copy of each processing step on the edge server, a copy on the site data center, and two copies on the cloud.

This approach presents several problems.
\begin{inparaenum}[(1)]
\item As operators are deployed on all nodes without considering the physical topology of the network, data communication is inefficient. For example, the instances of \texttt{FP} operators running in the cloud would need to collect data that could be efficiently filtered and preprocessed in a nearby edge server.
\item All operators are replicated on each machine, without considering the heterogeneity of resources. For instance, only cloud nodes may be equipped with suitable hardware to run the \texttt{ML} models.
\item The entire pipeline is instantiated as a single deployment unit, making it difficult to update the topology, e.g., if a new machine is added.
\end{inparaenum}

In summary, state-of-the-art dataflow systems lack \emph{locality awareness}, \emph{resource awareness}, and \emph{dynamic update mechanisms} to address the above scenario. In practice, in geographically distributed and heterogeneous scenarios like the one depicted above, engineers typically resort to manually deploying different parts of the processing pipelines on different sites, rather than using a single dataflow platform.
Starting from these premises, we propose a model that extends the dataflow programming and execution models to overcome the three limitations identified above, aiming to bring the ergonomics of the dataflow paradigm to geo-distributed and heterogeneous environments typical of the computing continuum.
\section{The FlowUnits Model}
\label{sec:model}

Moving from the motivations outlined in \s{background}, we propose a new programming and execution model to bring the benefits of dataflow to computing continuum scenarios.
Our model relaxes the assumption that computational hosts are physically co-located and homogeneous in terms of resources.
Instead, each host belongs to a geographical zone and is annotated with the hardware resources it provides, while programmers may declaratively specify where operators must run, how they must be replicated, and which resources they require.

\begin{figure}[tp]
\centerline{\includegraphics[width=1\columnwidth]{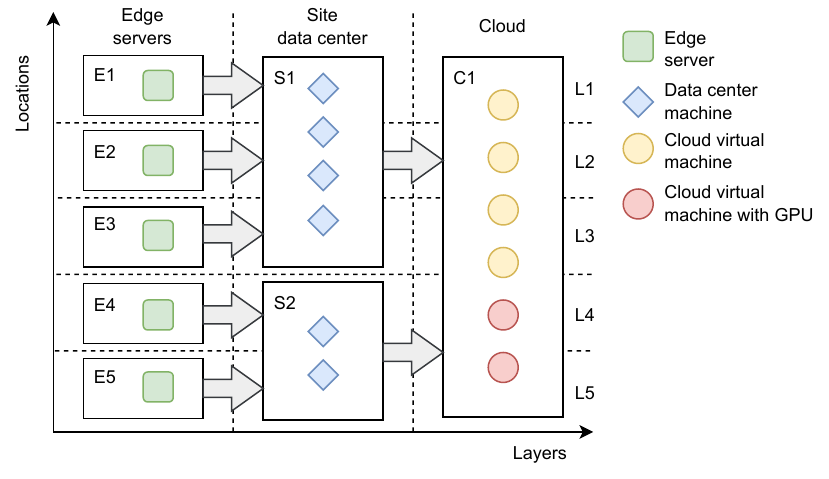}}
\caption{Organization of hosts into zones, identified by their layer and location.}
\label{fig:matrix}
\vspace{-2mm}
\end{figure}

%
\fakeparagraph{Geographical zones}
\fig{matrix} provides a graphical representation of our model, using again the example of the Acme company depicted in \s{background}.
Hosts are organized into geographical \emph{zones} (rectangles in \fig{matrix}), under the assumption that hosts within the same zone are well-connected to each other, allowing them to exchange data efficiently. In the example of the Acme company, a zone might represent a single edge server, or a site data center, or a cloud region.

Zones are defined in a two-dimensional space. Along one axis (x-axis in \fig{matrix}), zones are divided into \emph{layer}s that represent increasing levels of computational capabilities as we move from data sources at the periphery of the network (left in \fig{matrix}) toward the center (right in \fig{matrix}). Along the second axis (y-axis in \fig{matrix}), zones are divided according to their geographical \emph{location} (\texttt{L1} to \texttt{L5} in \fig{matrix}).

Zones are organized in a tree topology, which reflects the paths that data can follow when moving across zones. Zones belonging to the same level of the tree share the same layer (for example \texttt{edge}), but are in different locations.
\fig{matrix} represents the edges of the tree using gray arrows. It shows five edge servers, \texttt{E1} to \texttt{E5}, which are connected to the site data centers \texttt{S1} (the first three) and \texttt{S2} (the last two). Both site data centers are then connected to a single cloud zone \texttt{C1}.

As we will see later, communication between operators can only follow the path defined by the tree topology. For instance, hosts in \texttt{E1} and \texttt{E2} can both send data for processing to the \texttt{S1} data center, but they cannot send data to the \texttt{S2} data center.

%
\fakeparagraph{FlowUnits}
In our model, each operator of a dataflow graph is annotated with the layer where it must be executed. These annotations implicitly group operators into FlowUnits, such that contiguous operators in the dataflow graph that belong to the same layer are part of the same FlowUnit.
Considering the example in \fig{dataflow_example}, developers may specify that the filtering and preprocessing (\texttt{FP}) step must be executed on the edge, site anomaly detection (\texttt{AD}) must be executed in site data centers, and machine learning and analytics (\texttt{ML}) must be executed on the cloud. This would give rise to three FlowUnits, one for each processing step.

In turn, the entire computational job defined through a dataflow graph is annotated with the locations where it must be executed.
Returning to the example in \fig{matrix}, the job could be annotated with locations \texttt{L1}, \texttt{L2}, and \texttt{L4}.
This information, combined with the layer annotation that grouped operators into FlowUnits, will lead to instantiate a FlowUnit for each layer and for each location involved in the computation.
At the edge, there will be one instance of FlowUnit \texttt{FP} for each of the three edge servers \texttt{E1}, \texttt{E2}, and \texttt{E4}.
Among the site data centers, the infrastructure is organized into two zones: \texttt{S1} and \texttt{S2}. The first covers locations \texttt{L1} to \texttt{L3}, while the second covers locations \texttt{L4} and \texttt{L5}.
Consequently, for anomaly detection (\texttt{AD}) operators, which must be executed in site data centers, two FlowUnits will be instantiated: one for \texttt{S1}, which will receive data from edge servers \texttt{E1} and \texttt{E2}, and one for \texttt{S2}, which will receive data from edge server \texttt{E4}.
Similarly, operators part of FlowUnit \texttt{ML}, associated to the cloud layer, will be instantiated into the \texttt{C1} zone and will receive data from both \texttt{S1} and \texttt{S2} site data centers.

%
\fakeparagraph{Computational capabilities and requirements}
The presence of zones and FlowUnits extends the dataflow model by defining independent deployment units and making computation locality-aware. However, even within the same FlowUnit, some operators might have requirements that demand specific hardware (e.g., a GPU) and thus cannot be executed by every node within the zone where the FlowUnit is deployed.

To address this issue, in our model hosts are annotated with their \emph{capabilities} and operators are annotated with their \emph{requirements}. These annotations establish the mapping between dataflow operators and the physical infrastructure within a given zone.
Host capabilities are defined as attribute-value pairs that characterize the deployment targets. Each host in the network is associated with a set of capability descriptors that specify its hardware resources. These descriptors form a profile for each host that the runtime system uses during deployment. For example, a host might be annotated with:

\begin{itemize}
\item $n\_cpu = 8$ to indicate the machine has 8 CPU cores
\item $gpu = yes$ to specify GPU availability
\item $memory = 16GB$ to define available RAM
\end{itemize}

Operator requirements are expressed as logical conjunctions of Boolean predicates over the capability attributes. Each predicate specifies a constraint that must be satisfied by a host's capabilities for it to be considered a valid deployment target. The supported predicates include:

\begin{itemize}
\item Equality checks: $gpu = yes$ to require GPU presence
\item Numeric comparisons: $n\_cpu >= 4$ to require at least 4 CPU cores
\end{itemize}

A host satisfies an operator's requirements if and only if all predicates in the requirement specification evaluate to true when applied to the host's capabilities.

In our running example, the \texttt{ML} operator might require $n\_cpu >= 4 \wedge gpu = yes$, while in \fig{matrix} we see two types of cloud virtual machines, those that are equipped with a GPU (red circles) and those that are not (yellow circles). Assuming all machines in \texttt{C1} have more than 4 CPUs, the \texttt{ML} operator would be automatically deployed only on the red nodes of \texttt{C1}, while all other operators (not requiring a GPU) would be deployed on both red and yellow nodes.

%
\fakeparagraph{Dynamic updates}
Each FlowUnit is instantiated as a separate deployment unit, and our model admits multiple options for the communication between FlowUnits.
A supported option is the use of a persistent queuing system, which decouples the FlowUnits of different layers, simplifying management and updates.

When using a queuing system, adding a new geographical location only requires changing the annotation regarding which locations to replicate the computation on, without modifying anything else.
For example, if the computation was extended to location \texttt{L5}, the deployment would be updated accordingly, FlowUnit \texttt{FP} would be deployed to edge server \texttt{E5}, which would begin sending data to the queuing system from which machines in data center \texttt{S2} read, seamlessly integrating the new location into the existing processing pipeline.
Similarly, updating part of a computation pipeline can be done without disrupting the entire deployment.
For example, if the \texttt{ML} processing step was changed, only the \texttt{ML} FlowUnit should be interrupted and redeployed, while all other FlowUnits could continue to execute, pushing the results they produce or getting the data they need from the queuing system, even during \texttt{ML} update.

This flexibility enables real-time expansion or modification of the deployment topology as well as changes to the computation pipeline without service interruption, addressing the dynamic update challenge identified in \s{background}.

\section{Implementation}
\label{sec:impl}

We designed a prototype implementation of the proposed model on top of the Renoir dataflow system~\cite{renoir}. We selected Renoir because it is open-source and flexible~\cite{renoir_state}, and it offers a high level of performance~\cite{renoir_debs}.

Renoir exposes the classic dataflow model. The fundamental data structure in Renoir is a \textit{stream}, which represents a (possibly unbounded) sequence of elements, and offers a wide range of operators. Streams start from sources, for instance a file or a TCP connection, then are processed by operators that apply functional transformations to produce one or more output streams, and finally they are collected by sinks, for instance a file or a database, which store the final results of a graph of transformations. 
As discussed in \s{background}, a computation expressed as a logical dataflow graph is translated into a physical execution graph, where streams are divided into several partitions that the system can process concurrently.

The code snippet below exemplifies Renoir API using the classic world count example, where a stream of data is generated by reading from a file. Each line of the file is split into words (using the \texttt{flat\_map} operator). The resulting stream of words is partitioned into groups, where each group contains all instances of the same word (\texttt{group\_by} operator). A \texttt{fold} operator counts the number of occurrences within each group, and the final results (pairs of words with their associated count) are stored in a vector (\texttt{collect\_vec} operator).

\begin{lstlisting}[language=Rust,style=boxed]
let source = FileSource::new(path);
let result = context.stream(source)
    .flat_map(|line| line.split(" "))
    .group_by(|word| word.clone())
    .fold(0, |count, _word| *count += 1)
    .collect_vec();
\end{lstlisting}

The execution model of Renoir works as follows: a source file containing the specification of the job is compiled and executed on a single machine. Upon execution, Renoir reads the list of hosts to be involved in the computation from a configuration file and instantiates operators on each and every host using SSH.
Like most dataflow systems (see \s{background}), Renoir assumes all hosts to be physically co-located such that they can communicate efficiently. Consequently, it aims to maximize resource utilization by instantiating a copy of each operator in the dataflow graph for each CPU core made available by the hosts. Operator instances deployed on the same host communicate with in-memory channels, while operators in different hosts communicate through TCP connections.

Our implementation of the FlowUnits model in Renoir involves three changes.
\begin{inparaenum}[(1)]
\item We augment the API by adding two new methods to the \texttt{stream} data structure: \texttt{to\_layer}, which specifies that the subsequent chain of operators need to be deployed on a different layer of the edge-to-cloud continuum, and \texttt{add\_constraint}, which declares capability constraints for an operator.
\item We modify the deployment algorithm of Renoir, ensuring that operators are instantiated only on hosts belonging to the right layer and having sufficient capabilities to run them.
\item We introduce optional Kafka queues between FlowUnits, making sure that operators within each FlowUnit consume data and write data to the appropriate queues.
\end{inparaenum}
The code snippet below exemplifies the new API methods. The first part of the computation is performed on the \texttt{sensors} layer. It reads data from sensors, performs local filtering and data aggregation (a windowed average). Then, the computation is moved to the \texttt{cloud} layer, where the first \texttt{map} operator is executed on all the hosts available in that layer, whereas the second \texttt{map} operator is only executed on the hosts with capabilities that satisfy the given constraint.

\begin{lstlisting}[language=Rust,style=boxed]
let data = context
    .to_layer("sensors")
    .read_sensors()
    .filter(...).window().mean()
    .to_layer("cloud")
    .map(...)
    .map(...).add_constraint(...)
    .collect()
\end{lstlisting}

Our current prototype is intended as a proof of concept, and it is still not fully automated. Currently, the locations in which a computation should run and the association of layers and capabilities to hosts are written in a configuration file, which also provides the names of the queues used by FlowUnits to communicate.
We parse this configuration file using a Python script that generates the inventory file for the Ansible tool, which is responsible for the deployment of the executable.
\section{Evaluation}
\label{sec:eval} 

Our evaluation has two goals:
\begin{inparaenum}[(i)] 
\item assess the benefits of a topology-aware deployment, which exploits data locality; 
\item evaluate the effectiveness of our current implementation. 
\end{inparaenum} 

To do so, we simulate a distributed scenario that mimics the use case presented in \fig{matrix}. Specifically, we consider 4 edge servers (in 4 different zones) with low computational power (a single core each for simplicity), a site data center including 2 machines with 4 cores each, and a cloud platform including 1 virtual machine with 16 cores.
We run the experiments on a Workstation equipped with an AMD Ryzen 9 9950X, 64GB of DDR5 RAM, and a 1TB SSD storage, running Fedora Server 41 (kernel 6.13.5). 
%
To simulate the above scenario, we instantiated each node in a separate Docker container running Ubuntu 24.04. 
%
We used the \texttt{tc} (Traffic Control) Linux kernel utility to simulate varying network conditions between the edge, site, and cloud zones. We tested four different bandwidth limits: unlimited, 1 Gbit/s, 100 Mbit/s, and 10 Mbit/s. For each bandwidth setting, we also introduced three different network latencies: 0 ms (no latency added), 10 ms, and 100 ms. Connections within the same zone were assumed to have unlimited bandwidth and no added latency.

We defined a computational pipeline that consists of three operators: \texttt{O1}, \texttt{O2}, \texttt{O3}.
\texttt{O1} simulates initial data collection and preprocessing tasks by performing a simple computation that filters out 67\% of the data.
\texttt{O2} simulates intermediate data transformation and analysis by partitioning the input data, grouping it into windows and computing an average for each group.
\texttt{O3} implements an expensive processing task by computing the Collatz convergence steps for each item.

We compare two different deployment strategies.
The first one (denoted as \texttt{Renoir}) uses the standard execution strategy of Renoir, where a copy of each operator is instantiated for each core available in the entire deployment infrastructure. With the above configuration, Renoir instantiates 1 instance of each operator in each edge server, 8 instances of each operator in the site data center, and 16 instances of each operator in the cloud.
This is a standard approach, adopted also by other system such as Apache Flink~\cite{flink}, as it maximizes resource utilization when all hosts are co-located within the same cluster. However, this approach incurs a high communication overhead when data crosses slower links, as operators frequently exchange data across zone boundaries.
The second deployment strategy (denoted as \texttt{FlowUnits}) exploits the capabilities of our model to define locality-aware deployments, and instantiates \texttt{O1} only on the edge servers, \texttt{O2} only within the site data center, and \texttt{O3} only within the cloud. This minimizes inter-zone data transfer.
In this experiment, we are not considering dynamic updates to avoid measuring the overhead of an external queuing system, so FlowUnits communicate with each others through TCP connections.

\begin{figure}[tpb]
  \centerline{\includegraphics[width=\columnwidth]{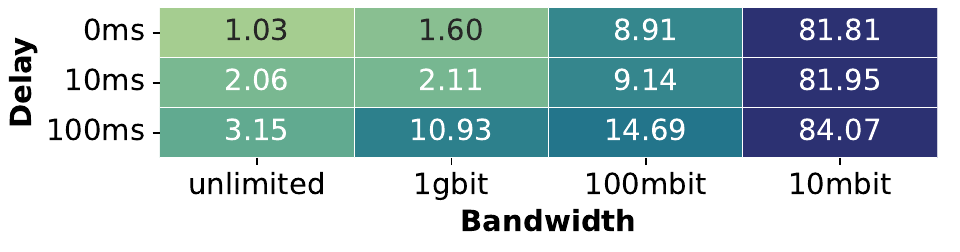}}
  \caption{Execution time ratio of a \texttt{Renoir} deployment vs a \texttt{FlowUnit} deployment, varying network bandwidth and delay between zones.}
  \label{fig:eval}
  \vspace{-2mm}
\end{figure}

For both deployments, we measure the overall execution time needed to process 10 million input events across the range of simulated network conditions.
\fig{eval} presents a heatmap summarizing the relative performance of the two strategies across all tested network conditions. The heatmap visualizes the ratio between the execution time of a \texttt{Renoir} deployment w.r.t. a locality-aware \texttt{FlowUnits} deployment (\texttt{Renoir}/\texttt{FlowUnits}). A ratio greater than 1 indicates that \texttt{FlowUnits} completed faster than \texttt{Renoir}.
As expected, the performance of the \texttt{Renoir} deployment is highly sensitive to network conditions. Its strategy of placing operator instances on all available cores leads to frequent data transfers between zones (edge-site, site-cloud). Consequently, when inter-zone network connections degrade, the communication overhead becomes a significant bottleneck, increasing the total execution time.
In contrast, the \texttt{FlowUnits} deployment, designed to minimize inter-zone communication, exhibits much greater resilience to network degradation.
These results validate the benefits of our topology-aware model and demonstrate the effectiveness of the \texttt{FlowUnits} approach in mitigating network bottlenecks in geographically distributed deployments.
\section{Related Work}
\label{sec:related}

The computing continuum paradigm has long been advocated as a solution to meet the challenges inherent in modern distributed applications, which are increasingly decentralized and need to timely handle large volumes of data~\cite{promise}.
However, consolidated data processing paradigms such as the widely used dataflow model still struggle to effectively support computing continuum scenarios.

In recent year, research has focused on addressing the challenges of \textit{locality-aware deployment}, \textit{heterogeneous resources}, and \textit{dynamic updates} in dataflow systems, but never in a unified way.
NebulaStream~\cite{nebula_1} is trying to optimize the dataflow execution on heterogeneous resources by using a query-on-demand mechanism to reduce the downstream load. The system also provides an automatic operator placement mechanism to optimize the dataflow execution. However, the system does not address the challenges of locality-aware deployment and dynamic updates.
DART~\cite{dart} builds a scalable stream processing system on top of a peer-to-peer distributed hash table. It deploys operators within the DHT considering both locality awareness and load balancing. It also focuses on fault-tolerance and adaptation to network changes by moving operators when nodes join or leave the network. Differently from our approach, it does not enable the definition of geographical zones that include well-connected nodes. It also does not address dynamic updates.
Pixida~\cite{pixida} aims to optimize data analytics across data centers. It proposes an algorithm that schedules tasks considering the cost for transferring data across data centers. While the approach focuses on a different execution model -- based on the scheduling of tasks rather than streaming of data across operators -- Pixida inspired the definition of geographical zones in our model.

Concerning dynamic updates, several solutions have been proposed.
Russo Russo et al.~\cite{hie_het_res} propose a hierarchical auto-scaling policy for data stream processing on heterogeneous resources. The system uses reinforcement learning to optimize the computation scaling at runtime. However, the system does not address the challenges of locality-aware deployment and dynamic updates.
Wang et al. present Fries~\cite{fries_dyn_rec}, a technique for performing runtime reconfigurations in dataflow systems with low delay, focussing on consistency and reducing delay during reconfiguration, solving part of the dynamic updates problem. However, the system does provide solutions to the challenges of locality-aware deployment and heterogeneous resources.
The interested reader can refer to a recent survey on dynamic adaptation mechanisms for data stream processing systems~\cite{adaptation}. In this work, the authors conclude that there still exists a gap in the research on dynamic adaptation for edge and computing continuum scenarios.

Our model was also inspired by multi-tier languages~\cite{multitier}, which enable specifying the location of data and computations and allow programmers to directly reference these concepts in their programs.
However, most existing languages target client-server architectures, and none of those presented in a recent survey focuses on a dataflow model and directly addresses the challenges addressed in this paper.

\section{Conclusions and Future Work}
\label{sec:conclusions}

In this paper, we have presented a new programming and execution model that extends the dataflow paradigm with locality awareness, resource awareness, and dynamic update capabilities, making it suitable for geo-distributed and heterogeneous edge-to-cloud computing environments.
Our FlowUnits model enables developers to write a complex processing pipeline once and deploy it across the computing continuum, transforming dataflow into an enabling technology for complex software architectures.
Future work will focus on enhancing our implementation to fully automate both initial deployment and subsequent updates, and evaluate it in real-world scenarios. At the model level, we are exploring formal correctness guarantees for dynamic updates to ensure data consistency and processing continuity across the continuum.

We believe that bridging dataflow programming with heterogeneous edge-to-cloud infrastructures represents a crucial advancement for enabling distributed data analysis and management in an increasingly decentralized computing landscape.
\section*{Acknowledgment}

\noindent
We acknowledge financial support from the PNRR MUR project PE0000021-E63C22002160007-NEST and the Inn. Grant COMET, CN00000013 National Centre for HPC, Big Data and Quantum Computing (HPC).

\bibliographystyle{IEEEtran}


\end{document}